\documentclass[aps,prb,reprint,superscriptaddress]{revtex4-1}

\usepackage{graphicx}
\usepackage{color}
\usepackage{amsmath}
\usepackage{bbm}
\usepackage{placeins}

\begin{document}

\title{Helical and topological phase detection\\ based on nonlocal conductance measurements in a three terminal junction.}
\author{P. W\'ojcik}
\email{pawel.wojcik@fis.agh.edu.pl}
\affiliation{AGH University of Science and Technology, Faculty of Physics and Applied Computer Science, al. A. Mickiewicza 30, 30-059 Krakow, Poland}
\author{D. Sticlet}
\affiliation{National  Institute  for  R$\&$D  of  Isotopic  and Molecular  Technologies,  67-103  Donat,  400293  Cluj-Napoca,
Romania}
\author{P. Szumniak}
\affiliation{AGH University of Science and Technology, Faculty of Physics and Applied Computer Science, al. A. Mickiewicza 30, 30-059 Krakow, Poland}
\author{M. P. Nowak}
\email{mpnowak@agh.edu.pl}
\affiliation{AGH  University  of  Science  and  Technology,  Academic  Centre  for  Materials
and  Nanotechnology,  al.   A.  Mickiewicza  30,  30-059  Krakow,  Poland}

\date{\today}

\begin{abstract}
The helical state is a fundamental prerequisite for many spintronics applications and Majorana zero mode engineering in nanoscopic semiconductors. Its existence in quasi-one-dimensional nanowires was predicted to be detectable as a characteristic reentrant behavior in the conductance, which in a typical two-terminal architecture may be difficult to distinguish from other possible phenomena such as Fabry-Perot oscillations.  Here we present an alternative method of helical gap detection free of the mentioned ambiguity, and based on the nonlocal conductance measurements in a three-terminal junction. We find that the interplay between the spin-orbit coupling and the perpendicular magnetic field leads to a spin-dependent trajectory of electrons and as a consequence a preferential injection of electrons in one of the arms. This causes a remarkable enhancement of nonlocal conductance in the helical gap regime. We show that this phenomenon can be also used to detect the topological superconducting phase when the junction is partially proximitized by an s-wave superconductor.
\end{abstract}

\maketitle
\section{Introduction}
Spin-orbit (SO) interaction, which couples the spin of electrons to their momentum, is an essential ingredient of many quantum devices including spin qubits~\cite{Nadjperge2010,Berg2013}, spintronic devices~\cite{Luo2011,Koo2009,Wojcik2014,Wojcik2017,Kohda2012,Debray2009}, Cooper pair splitters\cite{Hofstetter2009,Das2012} or Majorana nanowires\cite{Mourik2012,Albrecht2016, Sau2012}. 
In general, this relativistic phenomenon results from breaking the inversion symmetry which, in semiconductors,  could be either intrinsic, related to the crystallographic structure (Dresselhaus SO coupling\cite{Dresselhaus}), or may be induced by the confinement potential (Rashba SO coupling\cite{Rashba,Manchon2015}) in an electrically controllable fashion. The ability to control the Rashba SO component makes it especially important in spintronics applications and Majorana zero mode (MZM) engineering where the spin can be altered electrically via external gates attached to the nanostructure.\cite{Nadjperge2012,Wojcik2014,Wojcik2017,Nowack2007} For this reason, zincblende nanowires (NWs) grown along the $[111]$ direction, which preserves the crystal inversion symmetry and makes the Dresselhaus term negligible\cite{vanWeperen2015}, have attracted the growing interest in recent years\cite{vanWeperen2015,Campos2018,Iorio2019,Wojcik2019,Nadjperge2012,Wojcik2018,Wojcik2021}. 

In NWs characterized by large g-factor and strong Rashba SO interaction (InAs or InSb), a helical state may exist at finite magnetic fields, where electrons moving in opposite direction have opposite spins~\cite{Streda2003}.
This spin-momentum locking leads to a characteristic reentrant behavior in the conductance~\cite{Pershin2004} that in principle can be probed in a low-temperature transport measurement. Nevertheless, such measurements turned out to be challenging and a helical gap detection remained elusive for more than a decade. Up to date, the signature of the helical gap as a decline of conductance to 1$e^2/h$ at the 2$e^2/h$ plateau, with increasing magnetic field, has been reported for hole GaAs/AlGaAs wires\cite{Quay2010} as well as InAs gated nanowires\cite{Heedt2017}. Note however that even in those experiments, the appearance of conductance reentrance and its relation to the helical gap existence is not unambiguous and requires more sophisticated methods to distinguish it from other possible sources of the conductance drop such as Fabry-Perot oscillations~\cite{Rainis2014,vanWeperen2015}, Kondo effect~\cite{Heyder2015} or Coulomb interaction~\cite{Pedder2016}. More selective means for determination of the helical gap are devised in experiments which probe the conductance features when varying the magnetic field orientation~\cite{Kammhuber2017}.
Even in this case, however, as shown in Ref.~\onlinecite{Rainis2014}, the smoothness of the electrostatic potential profile between contacts and wire plays a crucial role and under some circumstances may mask the effects of SO interaction and the corresponding reentrant behavior of a conductance. 

Although the direct experimental measurement of the helical gap in NWs seems to be challenging, its presence has been probed indirectly by measurements of the signatures of MZM~\cite{Mourik2012}. Those end modes appear at the boundaries of a quasi-one-dimensional spinless p-wave superconductor, which is engineered by proximitizing a semiconductor NW in the helical state by an ordinary s-wave superconductor~\cite{Fu2008,Kitaev2003,Oreg2010,Alicea2012,Sarma2015}.

In this paper, we propose an alternative method of helical gap detection based on nonlocal conductance measurements in a three-terminal $T$-shaped junction. We find that in the presence of a magnetic field perpendicular to a three-terminal junction, the interplay between the spin-orbit coupling and the magnetic field results in the crosswise routing of electrons with opposite spins. In the helical state, when the injected electrons are spin-polarized, this leads to the injection preference to one of the arms which can be detected by a nonlocal conductance measurement.

Since the presence of the helical gap is a basic requirement for the topological superconducting phase and MZMs existence, we also demonstrate that the presented phenomenon can be used to detect the topological phase transition in the $T$-junction partially covered by a superconducting shell. The proposed method constitutes an alternative to local transport measurements that cannot distinguish trivial from topological bound states~\cite{PhysRevResearch.2.013377, pan2021quantized}. Superconducting/semiconducting multi-terminal hybrids have been recently fabricated either via lithography on 2DEGs~\cite{PhysRevLett.124.226801} or nanowire networks~\cite{Gazibegovic2017, Fadaly2017, PhysRevLett.125.116803} making our proposal viable within the current technology.

The  paper  is  organized  as  follows.  In  Sec.~\ref{sec:TheoreticalModel} the theoretical model and a description of the considered $T$-shaped junction are provided. Our proposal of the helical gap detection on the basis of a nonlocal conductance is given in~\ref{sec:results_a}, together with the explanation of the phenomenon standing behind it. In sec.~\ref{sec:results_b} we show how the proposed $T$-shaped nanostructure can be used for topological gap detection. Sec.~\ref{sec:summary} summarizes our results.

\section{Theoretical model}
\label{sec:TheoreticalModel}
\begin{figure}[!h]
	\begin{center}
		\includegraphics[scale=0.3]{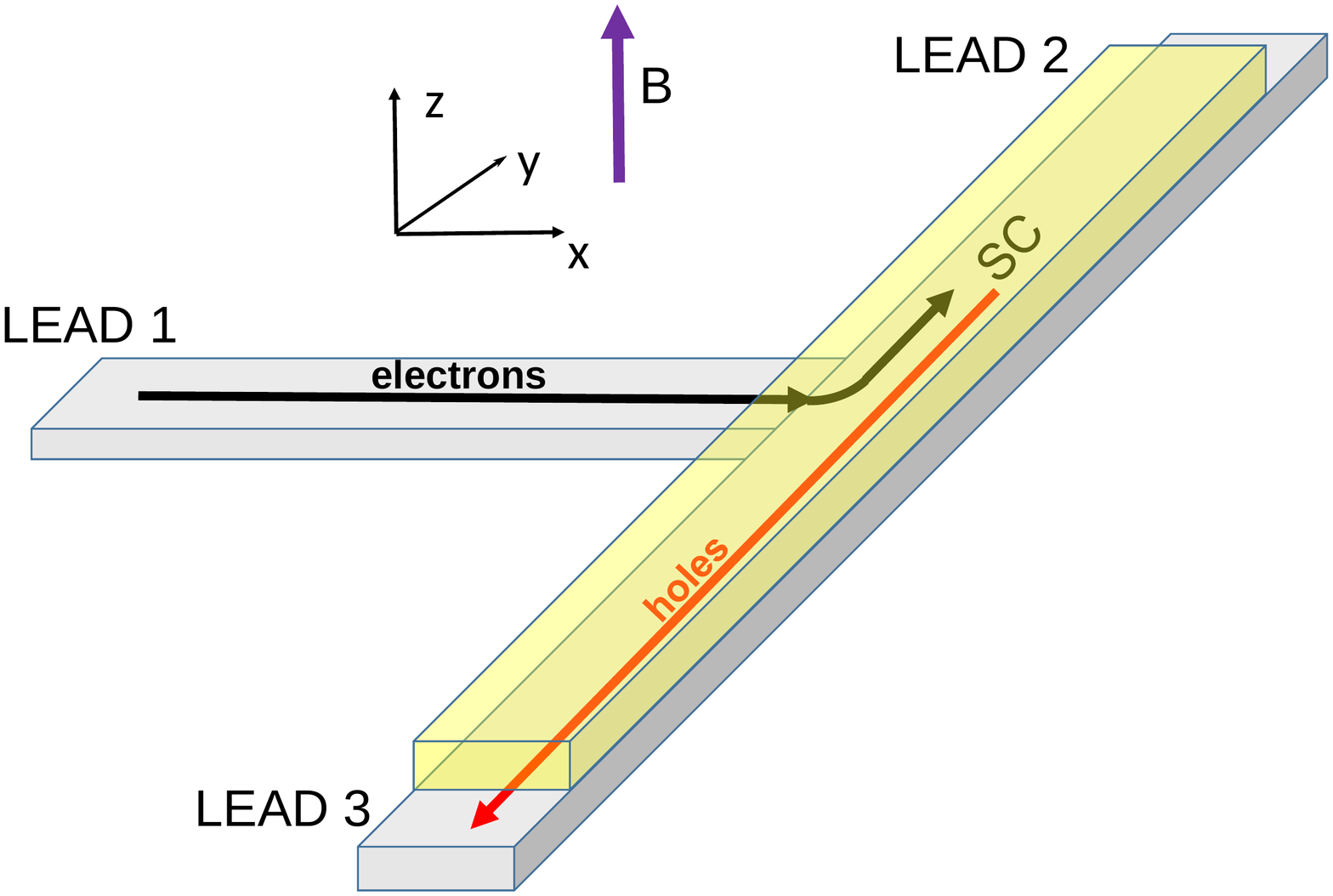}
		\caption{Sketch of the $T$-junction with one arm partially covered by a superconductor. Electrons are injected into the left lead $1$ and can flow out from the structure by all the leads $1,2,3$ as electrons or holes. The yellow shells correspond to a superconductor. In the sketch, the typical trace of charge in the helical gap regime is depicted for the configuration considered in sec.~\ref{sec:results_b}.
		}
		\label{fig1}
	\end{center}
\end{figure}

We consider a $T$-shaped device with quasi-one-dimensional leads with Rashba SO interaction and the superconducting pairing in one of the arms induced by a proximity to a superconductor, as depicted schematically in Fig.~\ref{fig1}.

In the presence of the magnetic field $\mathbf{B}=(0,0,B)$, the Hamiltonian of the system in the basis $(\hat{c}_{\mathbf{k}\uparrow}, \hat{c}^{\dagger}_{-\mathbf{k}\downarrow},\hat{c}_{\mathbf{k}\downarrow},-\hat{c}^{\dagger}_{-\mathbf{k}\uparrow})^{T}$ is given by
\begin{eqnarray}
\hat{H} &=& \left (\frac{\hbar ^2 \mathbf{k}^2}{2m^*} - \mu(\mathbf{r}) \right ) \sigma _0 \otimes \tau _z + \frac{1}{2}g\mu _B B \sigma_z \otimes \tau _0 \nonumber \\ 
&+& \alpha (\sigma _x k_y - \sigma _y k_x) \otimes \tau _z + \Delta(\mathbf{r}) \sigma _0 \otimes \tau _x,
\end{eqnarray}
where $\mathbf{k}=(k_x,k_y)$ with $k_{x(y)}=-i\hbar\: \partial / \partial x (y)$, $m^*$ is the effective mass, $\Delta(\mathbf{r}), \mu(\mathbf{r})$ are the spatially dependent pairing and chemical potential, $\sigma _i$ and $\tau _i$ with $i=x,y,z$ are the Pauli matrices acting on the spin and electron-hole degree of freedom, respectively. The dynamics of the electron spin is additionally determined by the Rashba SO coupling whose strength is defined by the parameter $\alpha$.

In our model, we adopt the  following  material parameters corresponding to  InSb:\cite{Gazibegovic2017} $m^*=0.014$, $g=-50$, $\alpha=50$~meVnm. For the device with induced superconductivity, we consider the induced gap $\Delta=0.2$~meV as obtained by proximitizing the semiconductor with a thin Al shell~\cite{Chang2015}. The numerical  calculations are performed on  a square lattice with $dx=dy=5$~nm and nanowire width $W=100$~nm. To determine the normal and Andreev transmission probabilities, we use the Kwant package~\cite{Groth2014}.

In the proposed nanostructure, electrons are injected  from the  left lead $1$, which acts as the input, and they either flow out from the device via the upper and lower arms or are reflected back into the input lead (see Fig.~\ref{fig1}).

At zero temperature, the nonlocal conductance $G_{ij}$ between the contacts $i$ and $j$ is given by\cite{Rosdahl2018,Anantram1996} 
\begin{equation}
G_{ij}= \frac{e^2}{h} \left (T^{ee}_{ij}-T^{he}_{ij}- \delta _{ij} N_i\right ),    
\label{eq:GSC}
\end{equation}
where $N_i$ is the number of transverse modes in the lead $i$ and $T^{ee}_{ij}, \: T^{he}_{ij}$ are the normal (electron-electron) and Andreev (electron-hole) transmission amplitudes corresponding to the situation when the electron is injected into the lead $j$ and flow out from the device as an electron or hole through the lead $i$.
For a non-superconducting system, the formula (\ref{eq:GSC}) reduces to the standard Landuer-B\"uttiker formula for a three terminal device~\cite{Wojcik2015} with $G_{ij}=\frac{e^2}{h}T_{ij}^{ee}$.

\section{Results}
\label{sec:results}
We shall now discuss the nonlocal conductance in the $T$-shaped nanostructure separately in the case with and without the superconducting shell. We put particular emphasis on the detection of the helical gap,
which results from the interplay between the Rashba SO coupling and the external magnetic field. 
We then discuss a possible detection of the topological phase by the nonlocal conductance measurement. 

\begin{figure}[!t]
	\begin{center}
		\includegraphics[scale=0.8]{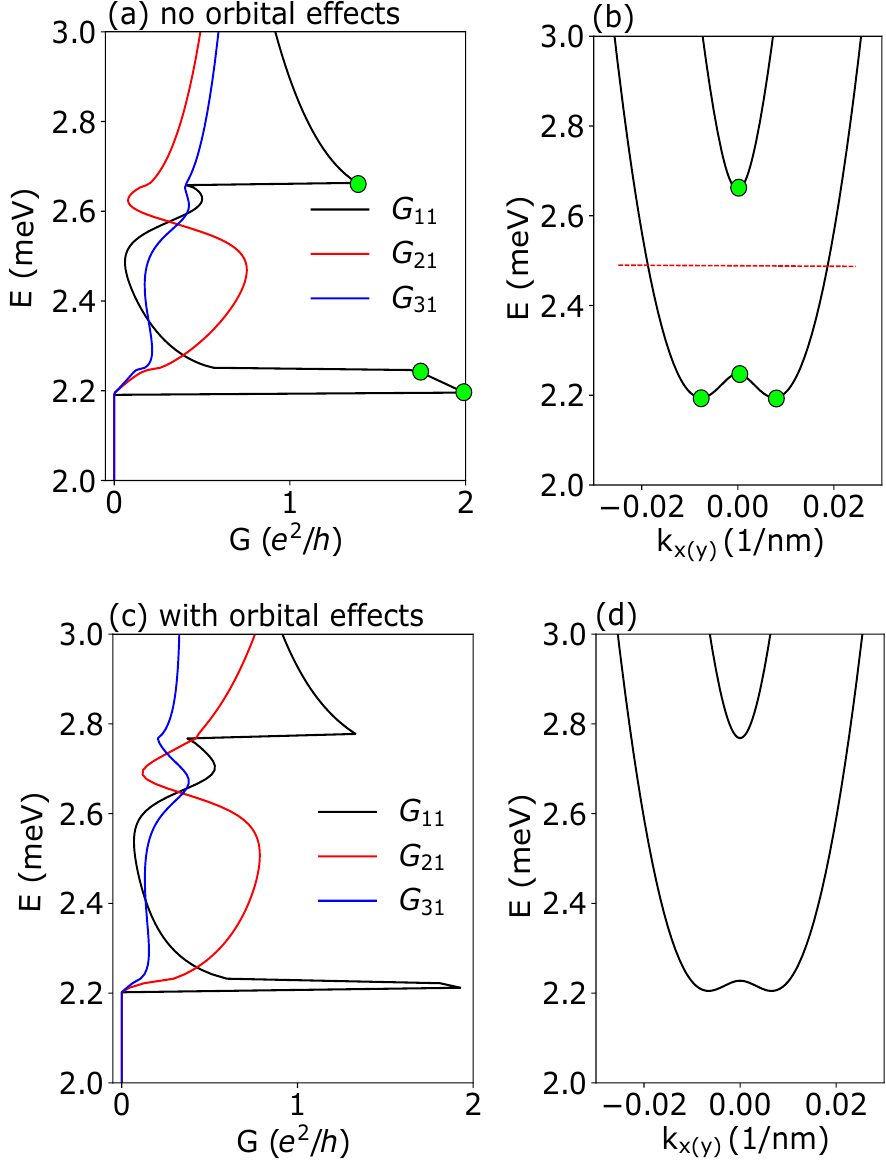}
		\caption{(a) Local $G_{11}$ and nonlocal $G_{21 (31)}$ conductance as a function the incoming electron energy $E$. In the helical gap energy range, $G_{21}$ (red line) is dominant due to the injection of electrons into the upper arm, preferred due to the interplay between the SO interaction and the Zeeman effect. (b) Dispersion relations (two lowest in energy subbands)  $E(k_{x(y)})$ in the horizontal (vertical) lead. Green dots in panel (b) correspond to the points in panel (a). Red dashed line marks the value of energy $E$  chosen for a further analysis. Panels (c) and (d) presents the same as (a) and (b) correspondingly, but with the inclusion of the orbital effects of the magnetic field. We can see that orbital effects does not affect significantly the results. Here, we set $\mu=0$, $B=0.15$~T. }
	 \label{fig2}
	\end{center}
\end{figure}

\subsection{Helical gap detection}
\label{sec:results_a}
\begin{figure*}[!ht]
	\begin{center}
		\includegraphics[scale=0.2]{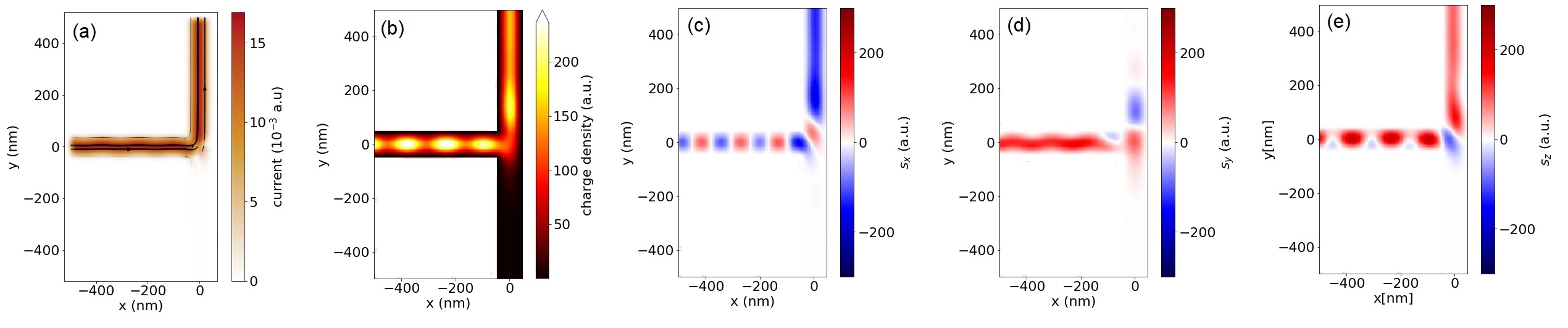}
		\caption{(a) Current and (b) electron density in the $T$-junction with a clear preference of injection into the upper arm. Panels (c)-(e) present the spin densities $s_x$, $s_y$ and $s_z$, respectively. Results for $E=2.45$~meV and $B=0.15$~T.}
		\label{fig3}
	\end{center}
\end{figure*}

Let us first consider the $T$-shaped junction as in Fig.~\ref{fig1} without the superconducting shell. For the considered nanostructure we set $\mu=0$ and calculate the nonlocal conductance as a function of the incoming electron energy [see Fig.~\ref{fig2}(a)]. 
In nanowires with strong Rashba SO coupling, the spin-Zeeman effect of the perpendicular magnetic field opens the helical gap. We illustrate this by plotting the dispersion relation $E(k_{x(y)})$ for the horizontal (vertical) system lead as presented in Fig.~\ref{fig2}(b). In this specific energy range, the spin of an electron is directly coupled to its momentum, so that in particular for one-dimensional nanowires, electrons flowing in opposite directions possess opposite spins.  As seen in  Fig.~\ref{fig2}(a), for the energy range corresponding to the helical gap [compared with panel (b)], the nonlocal conductance $G_{21}$ dominates over other conductance components. This is a consequence of the preferred electron injection to the upper arm, clearly seen on the current map in Fig.~\ref{fig3}(a). Note however, that the preference does not result from the orbital effects of the magnetic field and the corresponding Lorentz force, which are not included here. Results with the inclusion of the orbital effects\cite{orbital} are presented below in panels Fig. ~\ref{fig2}(c,d), and do not exhibit any significant difference with respect to those depicted in panels (a),(b). For this reason,  henceforth we neglect the orbital effects of the magnetic field. 

The injection preference presented in Fig.~\ref{fig3} is a consequence of  the combined effects of the Rashba SO coupling and the magnetic field and can be explained based on the Heisenberg equation, according to which the second derivative of the position operator $\mathbf{\hat{r}}$, classically interpreted as a force, can be expressed as 
\begin{eqnarray}
    \mathbf{\hat{F}}(t)&=&m^*\frac{d^2\mathbf{\hat{r}}}{dt^2}= \frac{m^*}{\hbar^2} [\hat{H},[\mathbf{\hat{r}},\hat{H}]] \nonumber \\ &=&\frac{m^*}{\hbar^2}  \left \{ 2 \alpha ^2 (k_y,-k_x) \sigma_z - g\mu_B B_z \alpha (\sigma _x, \sigma _y) \right \}.
    \label{eq:F}
\end{eqnarray}
Since the force operator $\mathbf{\hat{F}}$ depends on the electron spin, which is an internal quantum degree of freedom, it does not have a classical analog. 
The first term is related to the well-known internal spin Hall effect\cite{Sinova2004} whereas the second corresponds to the interplay of the perpendicular magnetic field with the Rashba SO coupling.  The physical meaning, i.e., the measurable prediction is contained in the quantum-mechanical expectation value, defined as $\langle \hat{\mathbf{F}} (t) \rangle = \langle \Psi (t=0) |\hat{\mathbf{F}}(t)| \Psi(t=0) \rangle $, with the initial spinor $\Psi (t=0)=(\psi^{\uparrow}_e(t=0), \psi^{\downarrow}_e(t=0))^T$.

\begin{figure}[!h]
	\begin{center}
		\includegraphics[scale=0.35]{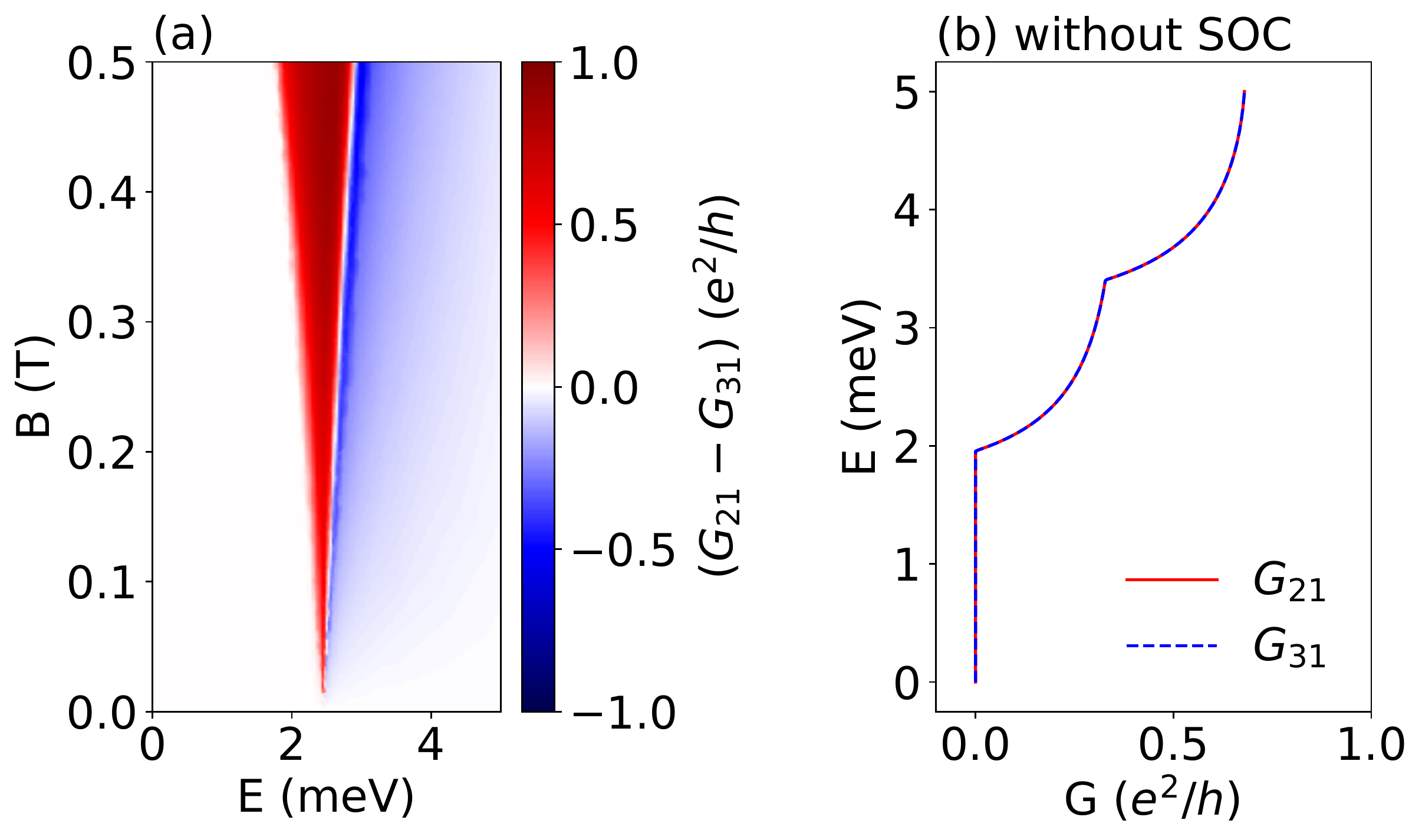}
		\caption{(a) Nonlocal conductance asymmetry  defined as $G_{21}-G_{31}$ vs. energy of the incoming electron and the magnetic field $B$. Panel (b) presents the nonlocal conductance $G_{21}(E)$ and $G_{31}(E)$ calculated without SO coupling at $B=0.3$~T.}
		\label{fig4}
	\end{center}
\end{figure}

For the sake of simplicity, let us now reduce the system to one-dimensional nanowires connected in the $T$-shape geometry. We assume that the electron with energy $E=0$ from the helical gap range is injected into the input lead $1$ within the quantum state corresponding to $+k_x$,
\begin{small}
\begin{equation}
    \Psi^{1D}_{+k_x}=\frac{1}{\sqrt{2}(\Delta _Z^2 + \alpha ^2 k_x^2)^{\frac{1}{4}}}
    \left (
    \begin{array}{c}
         \sqrt{(\Delta _Z^2+\alpha ^2 k_x^2)^{\frac{1}{2}}-\Delta _Z}  \\
         \sqrt{(\Delta _Z^2+\alpha ^2 k_x^2)^{\frac{1}{2}}+\Delta _Z}
    \end{array}
    \right )e^{ik_xx},
\end{equation}
\end{small}

\noindent where $\Delta _Z=\frac{1}{2}g\mu_B B$ is the Zeeman energy. It flows through the horizontal NW with the force expectation value $\langle \hat{\mathbf{F}} (t) \rangle =0$, characteristic for all eigenstates. At the fork of the $T$-junction, the electron in the state $\Psi^{1D}_{+k_x}(t=0)$ is injected into the vertical NW for which $\Psi^{1D}_{+k_x}(t=0)$ is not an eigenstate. As a result, $\langle \hat{\mathbf{F}} (t) \rangle$ is no longer zero and takes the form
\begin{eqnarray}
\langle \hat{\mathbf{F}} (t) \rangle = -\frac{2m^*\Delta_Z \alpha ^2}{\hbar ^2 \sqrt{\Delta _Z^2 + \alpha ^2 k_x ^2}}(k_y,k_x)
\label{eq:force}
\end{eqnarray}
with the non-zero $y$ component pushing electrons into the upper arm (note that $\Delta_Z<0$ as we assume $g=-50$).
Although $\langle \hat{\mathbf{F}} (t) \rangle _y $ increases linearly with the magnetic field, it is non-zero only when the SO coupling is present. Indeed, as shown in Fig.~\ref{fig4}(b), for $\alpha=0$ and $B\ne 0$ electrons are injected symmetrically in both the junction arms leading to the equal non-local conductances  $G_{21}$ and $G_{31}$. 

The strong injection preference is characteristic for the energy range corresponding to the helical gap when the electron in the helical state is injected into the system with a single well-defined $k$ vector. In this energy range, the backscattering requires the spin flip, which is energetically less favorable than nearly half spin rotation needed for the injection into the upper arm -- see Fig.~\ref{fig3}. This, apart from the SO induced force (\ref{eq:force}), additionally enhances the observed electron injection preference.

For energies outside the helical gap range, electrons can flow in the horizontal NW within two states defined by different wave vectors, characterized by different spin orientations. As such, they are injected into the opposite $T$-junction arms or are backscattered. As a result, the dominance of the $G_{21}$ conductance is observed solely within the parameter regime ($B$, $E$) corresponding to the helical gap. 

The magnetic field dependence of the nonlocal conductance asymmetry defined as $G_{21}-G_{31}$ is presented in Fig.~\ref{fig4}(a) and clearly indicates the parameter range where the helical gap occurs.

\subsection{Topological phase detection}
\label{sec:results_b}

As the nonlocal conductance measurement in the $T$-shaped junction clearly indicates the helical gap range, we now analyze if the same effect can be used to detect the topological superconducting phase which requires the helical gap existence\cite{Oreg2010}. For this purpose, we consider the system depicted in Fig.~\ref{fig1} with the superconducting shell covering the fork. The superconducting shell induces electron pairing underneath, in the semiconductor nanowire, and under appropriate conditions the system undergoes the topological phase transition resulting in zero-energy Majorana states localised at the ends of the superconducting section. In this configuration, electrons from the horizontal nanowire are injected directly into the superconducting vertical arm where they can be reflected as holes or transmitted to the leads $2$ and $3$ as holes or electrons. We assume that the length of the superconducting shell $L_{SC} < \xi$, where $\xi$ is the superconducting coherence length. Note that if the chemical potential is situated in the helical gap in the superconducting nanowire, needed for a topological phase transition, we should observe the nonlocal conductance asymmetry in analogy to Fig.~\ref{fig4}(a).

\begin{figure}[!h]
	\begin{center}
		\includegraphics[scale=0.4]{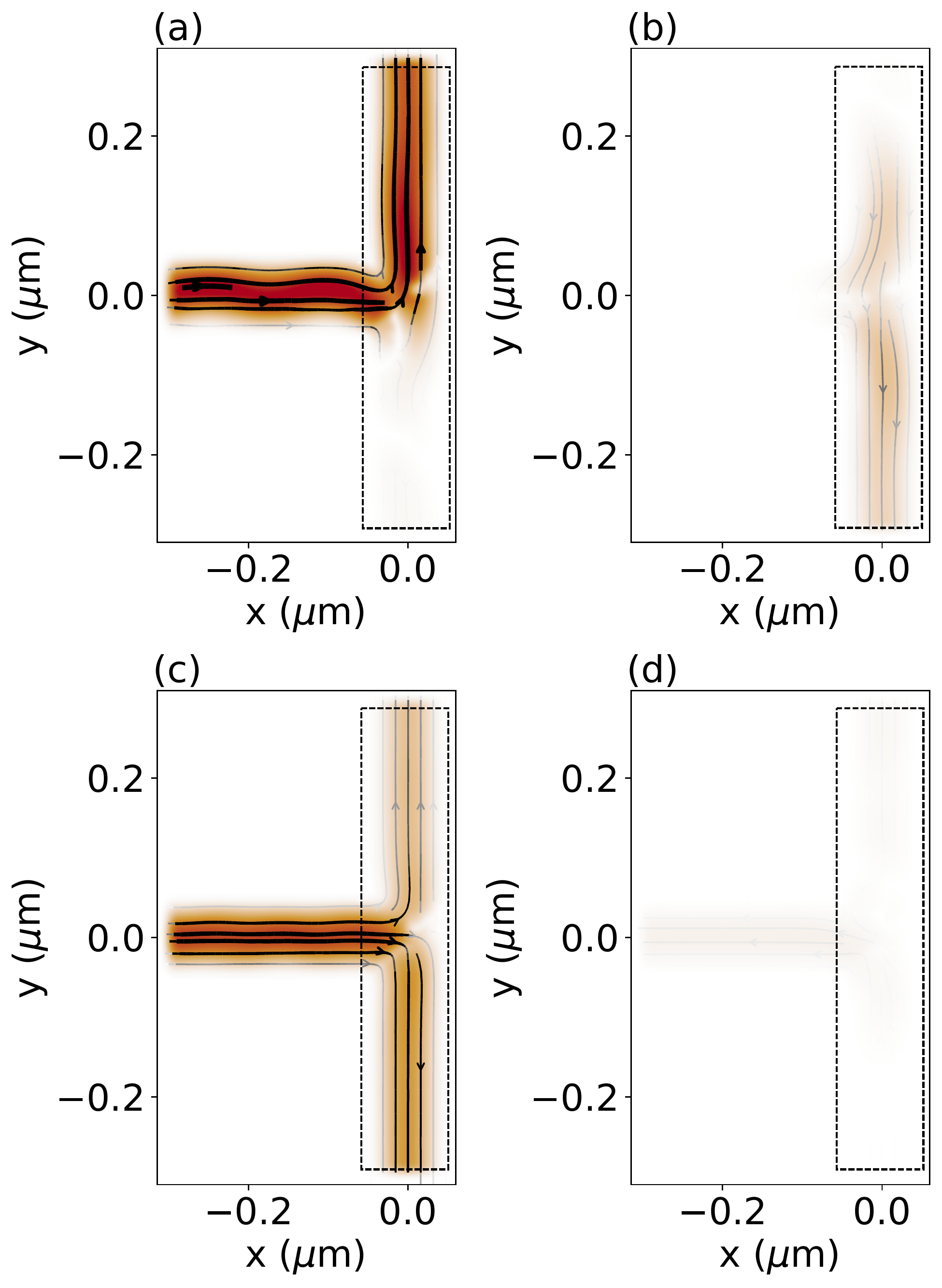}
		\caption{(a)(c) Electron and (b)(d) hole current density for $B=0.4$~T. In (a)(b) the chemical potential in the whole system is set to $\mu=2.45$~meV such the normal leads are in the helical regime, while the proximitized region is in the topological state. In (c)(d) $\mu=4$~meV when the proximitized region is in the trivial range and there is no helical state in normal leads. Dashed rectangles mark the superconducting shell.}
		\label{fig5}
	\end{center}
\end{figure}

\begin{figure}[!h]
	\begin{center}
		\includegraphics[scale=0.35]{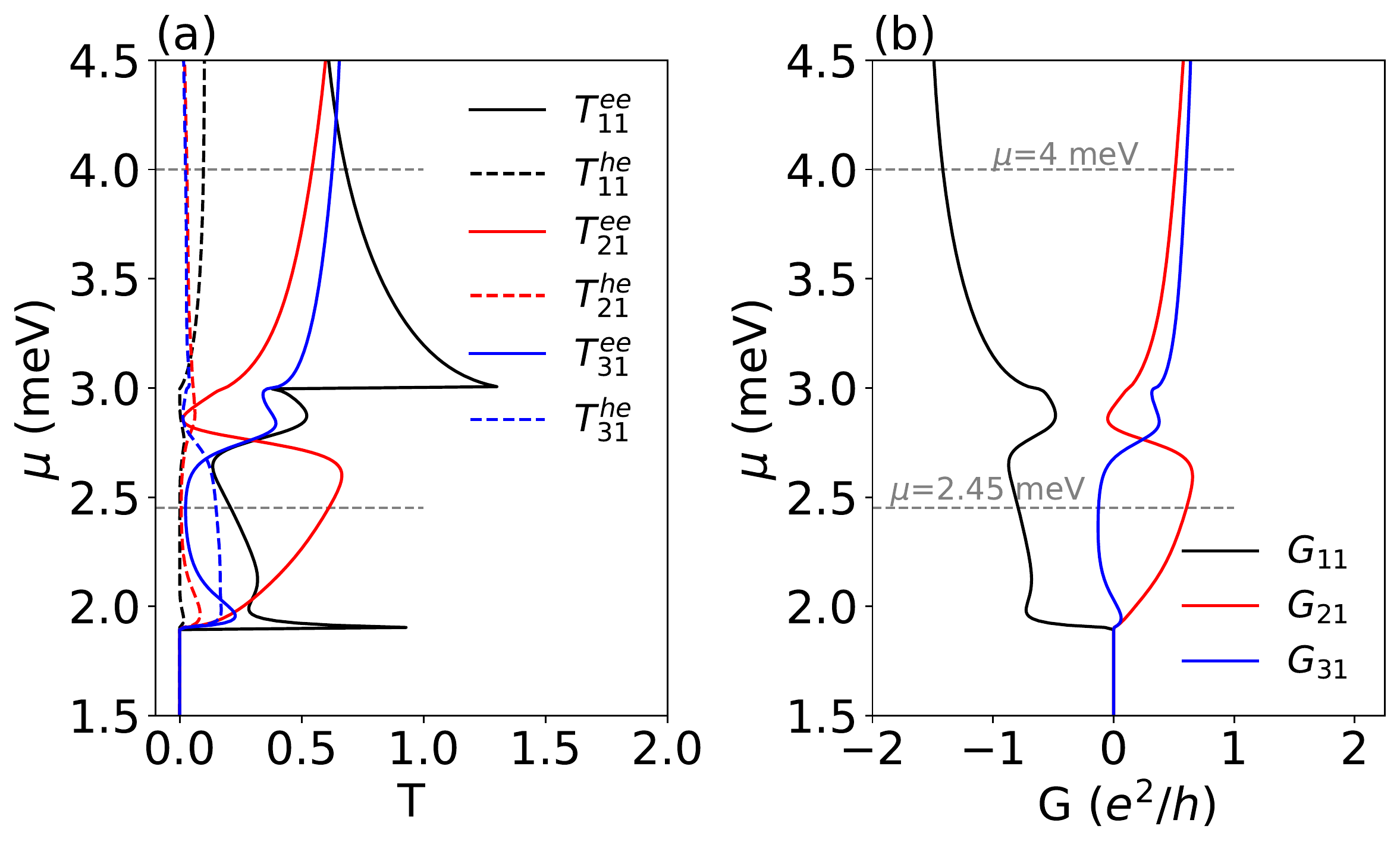}
		\caption{(a) Zero energy transmission coefficients and (b) the local and nonlocal conductance as a function of the chemical potential for the $T$-shaped junction partially proximitized by a superconductor. Results for $B=0.4$~T. Gray horizontal lines present values of $\mu$ chosen for the analysis -- see Fig.~\ref{fig5}.}
		\label{fig6}
	\end{center}
\end{figure}

As previously, the electron is injected into the nanostructure from the lead $1$. If its energy lies in the helical gap at the fork, it turns left into the upper arm due to the mechanism described above. Here, if $E<\Delta$ it undergoes the Andreev reflection as a hole with opposite spin [cf. Fig.~\ref{fig5}]. The reflected hole is transmitted to lead $3$. Note that due to the finite length of the proximitized region, the Andreev reflection is not perfect, and there is a finite probability for an electron to escape through the lead $2$. This process results in nonzero values of the transmission amplitudes $T_{21}^{ee}$, $T_{31}^{he}$ (and $T_{11}^{ee}$ due to the electron reflection from the fork) as can be seen around $\mu=2.45$~meV in Fig.~\ref{fig6}(a). This in turn results in the pronounced difference of the nonlocal conductances $G_{21}$ and $G_{31}$ as presented in Fig.~\ref{fig6}(b).

On the contrary, when the system is in a trivial regime (i.e., there is no helical gap in the normal leads and no topological gap in the proximitized region) obtained for $\mu=4$~meV, the electron current is distributed almost symmetrically among leads 2 and 3. The Andreev reflection is marginal as can be seen in Fig.~\ref{fig5}(c) and (d). As a result, the corresponding nonlocal conductances $G_{21}$ and $G_{31}$ take similar values depicted in Fig.~\ref{fig6}(b). The above shows that the directional electron flow is inherited by the superconducting system provided that it is in the topological superconductivity phase.

So far, we assumed that the $T$-junction can be regarded as a homogeneous system with a constant chemical potential across the nanostructure. The electron injected into the horizontal nanowire within the helical state remains in this helical gap energy range also when it flows through the vertical arm. Note however, that in practice, the assumption of the constant $\mu$ throughout the nanotructure is difficult to fulfill experimentally, as the electronic structure of the vertical nanowire can be affected by the presence of the superconducting shell.

\begin{figure}[!h]
	\begin{center}
		\includegraphics[scale=0.35]{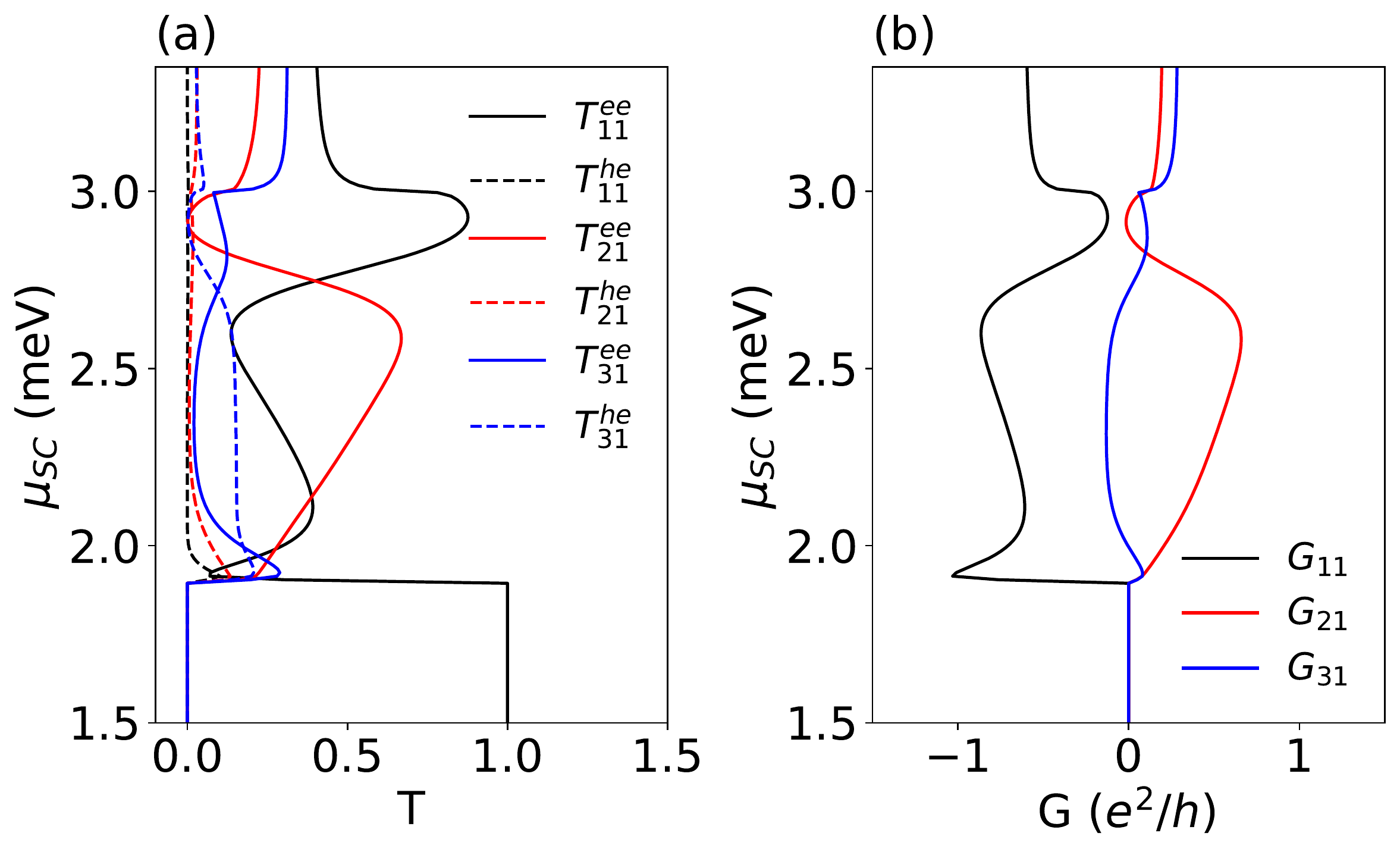}
		\caption{(a) Zero-energy transmission coefficients and (b) the local and nonlocal conductance as a function of the chemical potential $\mu_{\mathrm{SC}}$ in the proximitized region while in the rest of the system it is kept constant at $\mu=2.45$~meV. Results for $B=0.4$~T.}
		\label{fig7}
	\end{center}
\end{figure}

In Fig.~\ref{fig7}(a) we present the transmission coefficients for a system where in the normal part we keep a constant value of the chemical potential $\mu=2.45$~meV and vary only the chemical potential $\mu_{\mathrm{SC}}$ in the superconducting region. In the range of $\mu_{\mathrm{SC}}$, in which the proximitized part is in topological phase, we again observe  pronounced $T_{21}^{ee}$, $T_{31}^{he}$ amplitudes which results in significant differences in $G_{21}$ and $G_{31}$ as presented in Fig.~\ref{fig7}(b).

\begin{figure}[!h]
	\begin{center}
		\includegraphics[scale=0.6]{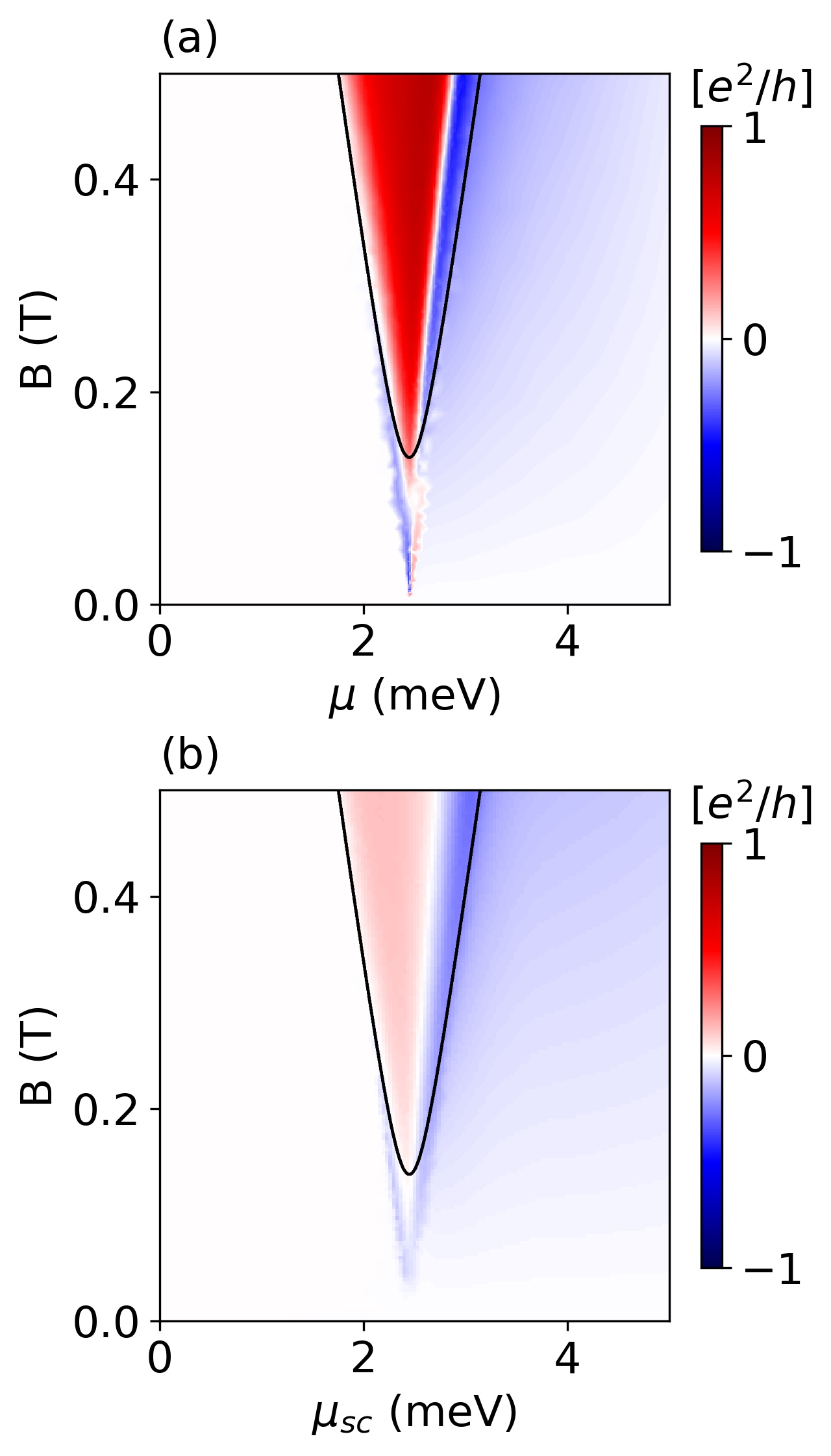}
		\caption{Zero-energy nonlocal conductance asymmetry $G_{21}-G_{31}$ as a function of the chemical potential $\mu$ and the magnetic field $B$. Results for the case (a) when the chemical potential is assumed to be constant throughout the nanostructure and (b) when the chemical potential in the horizontal nanowire is fixed at the energy $\mu=4$~meV, beyond the helical gap regime. Black lines denote the topological phase transition calculated for a single nanowire with parameters corresponding to the superconducting part of the $T$-junction.}
		\label{fig8}
	\end{center}
\end{figure}

In Fig.~\ref{fig8}, the asymmetry of the zero energy nonlocal conductance $G_{21}-G_{31}$ is clearly apparent in the case (a) when the chemical potential is assumed to be constant throughout the nanostructure or even when (b) the chemical potential in the horizontal nanowire is fixed beyond the helical gap at the energy $\mu=4$~meV. Note that in the latter case, there are two spin bands available in the input lead $1$. As predicted from the analytical model, Eq.~(\ref{eq:force}), the expectation value of the force depends on $k_x$ at which the electron is injected into the vertical nanowire, which is different for two spin-orbit split bands preserving the asymmetry of the nonlocal conductances $G_{21}$ and $G_{31}$.

\begin{figure}[!h]
	\begin{center}
		\includegraphics[scale=0.5]{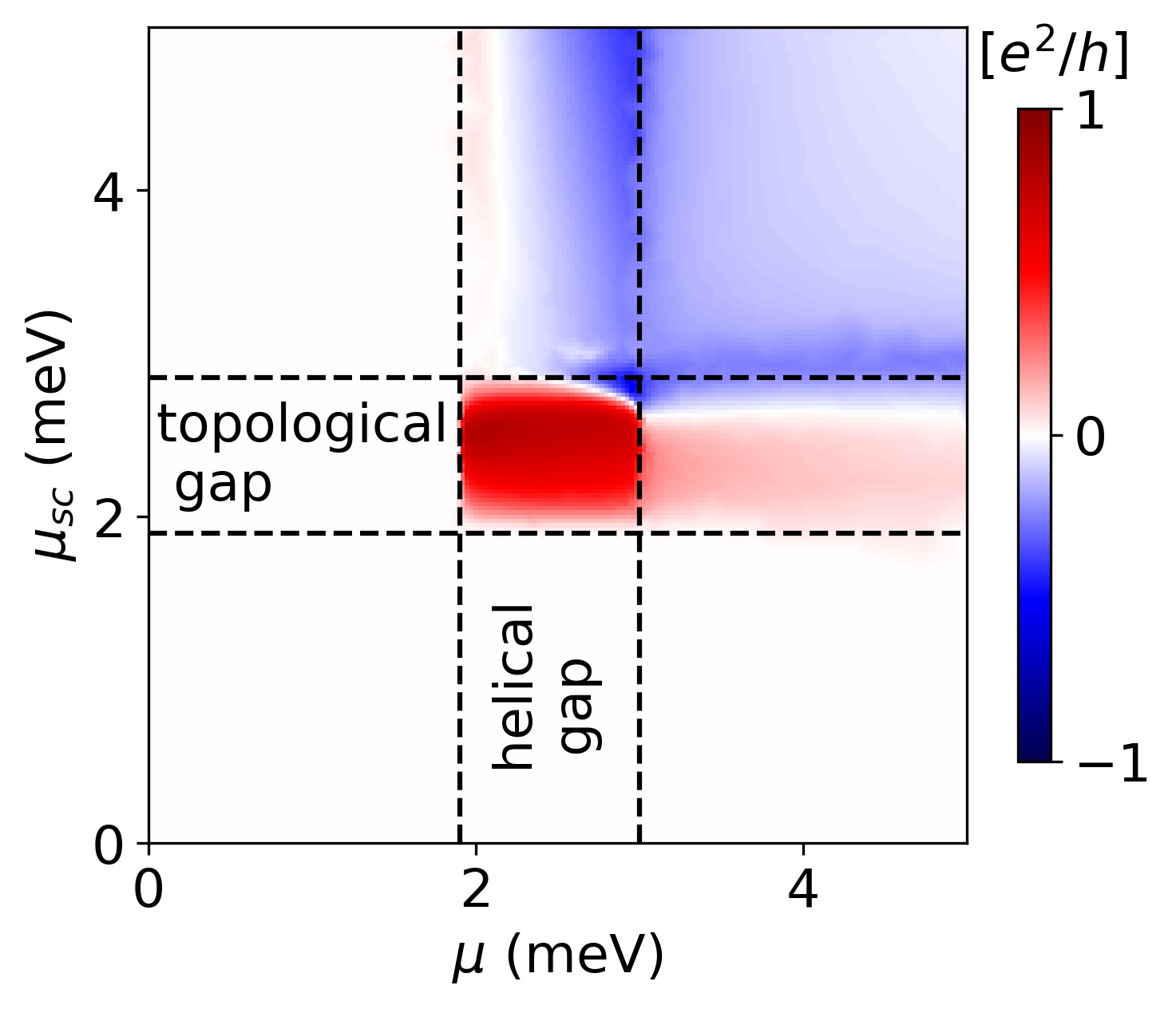}
		\caption{Nonlocal conductance asymmetry $G_{21}-G_{31}$ as a function of the chemical potential in the input horizontal nanowire $\mu$ and in the superconducting vertical nanowire $\mu_{\mathrm{SC}}$. Results for $B=0.4$~T. }
		\label{fig9}
	\end{center}
\end{figure}

In Fig.~\ref{fig9} we present the asymmetry of the zero-energy nonlocal conductance $G_{21}-G_{31}$ calculated for a fixed $B=0.4$~T as a function of the chemical potential in the input nanowire $\mu$ and in the superconducting vertical nanowire $\mu_{\mathrm{SC}}$. The helical gap range in the input nanowire and the topological gap range in the superconducting nanowire are marked by dashed vertical and horizontal lines, respectively. 

As we see, the difference in nonlocal conductance is a clear hallmark of the presence of a topological phase. The conductance asymmetry is obtained regardless of the chemical potential in the horizontal lead and even if the magnetic field is switched off in the horizontal nanowire, preventing helical gap creation in it. 

\section{Summary and conclusions}
\label{sec:summary}
Using the scattering matrix approach, we have analyzed transport properties of a $T$-shaped junction characterized by the strong SO coupling. In the presence of a perpendicular magnetic field, we have found that electrons injected with energy in the helical gap regime are turned into the upper arm leading to the remarkable enhancement of the nonlocal conductance. This behavior has been explained within the Heisenberg equation as resulting from the existence of a spin-dependent force which results from the interplay between the SO interaction and the magnetic field. The force pushes electrons with opposite spins into the opposite sides of the nanowire. 
Specifically, in the helical gap range, when the electron spin is bound to its momentum, the electrons are preferably injected into one of the arms of the $T$-junction. This, as proposed, allows for detection of the helical gap regime by employing nonlocal conductance measurements. As we noticed, the proposed method is free of possible misinterpretations resulting from Fabry-Perrot oscillations present in the two-terminal architecture and local conductance measurements. 

Next, by partially covering the $T$-junction by a superconducting shell, we analyze if the same phenomenon can be useful in topological phase detection. We have shown that the nonlocal conductance measurements in the $T$-shaped superconducting junction clearly indicate the topological phase regime even if the chemical potential in both the vertical and horizontal nanowire is not perfectly adjusted. 

Note that the calculations presented in the paper do not include the orbital effects. Although in sec.~\ref{sec:results_a} we have shown that for the considered magnetic field the orbital effects do not play an important role, the investigation of their influence on the superconducting shell and the topological phase is beyond the scope of this paper and was presented in detail in one of our previous papers.\cite{Nowak2018,WojcikNowak} 

As a final remark, note that the theoretical predictions presented in this paper should be verifiable within the current state of the art epitaxial technology, which allows for the fabrication of high-quality $T$-shaped, $X$-shaped or hashtag nanowire networks (also with superconducting shells) and performing precise conductance measurements\cite{Gazibegovic2017, Fadaly2017}.

\section{Acknowledgement}
This  work  was  supported  by  National  Science  Centre (NCN) decision number DEC-2016/23/D/ST3/00394 and by the Romanian Ministry of Research, Innovation and Digitization, CNCS/CCCDI – UEFISCDI, project number PN-III-P1-1.1-TE-2019-0423, within PNCDI III, and partially by PL-Grid and TeraACMiN Infrastructure.

\begin{thebibliography}{52}
\bibitem [{orb()}]{orbital}%
  \BibitemOpen
  \href@noop {} {\ }\bibinfo {note} {The orbital effects of the magnetic field
  are incorporated using Peierls substitution of the hopping elements $t_{nm}
  \rightarrow t_{nm} \exp {\left [ -ie \int \mathbf{A}d\mathbf{l} /\hbar\right
  ] }$}\BibitemShut {NoStop}%
\end{thebibliography}

%

\end{document}